\documentclass[prl,twocolumn]{revtex4}

\input{epsf}

\usepackage{graphicx}
\usepackage{dcolumn}
\usepackage{bm}

\usepackage{amsmath}

\begin{document}

\title{Formation of $\eta'(958)$-mesic nuclei and axial $U_A(1)$ anomaly
at finite density}

\author{Hideko Nagahiro}
\author{Satoru Hirenzaki}
\affiliation{%
Department of Physics, Nara Women's University, Nara
630-8506, Japan
}%

\date{\today}

\begin{abstract}
We discuss the possibility to produce the bound states of the $\eta'(958)$
 meson in nuclei theoretically. We calculate the formation cross sections
 of the $\eta'$ bound states with the Green function method for ($\gamma$,p)
 reaction and discuss the experimental feasibility at photon facilities
 like SPring-8. We conclude that 
we can expect to observe resonance peaks in ($\gamma$,p) spectra for the
 formation of $\eta'$ bound states and we can deduce new
 information on $\eta'$ properties at finite density. These
 observations are believed to be essential to know the possible mass
 shift of 
 $\eta'$ and deduce new information of the effective restoration of the
 chiral $U_A(1)$ anomaly in the nuclear medium.
\end{abstract}

\maketitle

In the contemporary hadron physics,
the light pseudoscalar mesons ($\pi$, K, $\eta$) are recognized as the
Nambu-Goldstone bosons associated with the spontaneous breaking of the
QCD chiral symmetry.
In real world, these mesons,
together with heavier $\eta'(958)$ meson,
 show the involved mass spectrum, which are
believed to be explained by the explicit flavor 
$SU(3)$ breaking due to current quark masses and the breaking of the
axial $U_A(1)$ symmetry at the quantum level referred as the $U_A(1)$
anomaly~\cite{PLB206,NPB308}.
One of the most important subjects in hadron physics at present is to
reveal the 
origin of the hadron mass spectra and to find out the quantitative
description of hadron physics from QCD~\cite{PR247}.

Recently, 
there are several very important developments for the study of the
spontaneous breaking of chiral symmetry and its partial restoration at
finite density.
To investigate
the in-medium behavior of spontaneous chiral symmetry breaking,
the hadronic systems, such as pionic atoms~\cite{PLB514etc,PRL92,
APPB31etc}, $\eta$-mesic 
nuclei~\cite{our_eta,valencia_eta} and $\omega$-mesic
nuclei~\cite{omega}, have been 
investigated in both of theoretical and experimental aspects.
Especially,
after a series of deeply bound pionic atom experiments~\cite{PRC62etc,PRL88},
K.~Suzuki {\it et al.} reported the quantitative
determination of pion decay constant $f_\pi$ in-medium from the deeply
bound pionic states in Sn isotopes~\cite{PRL92} and stimulated many active researches
of the partial restoration of chiral symmetry at finite
density~\cite{PLB514etc,APPB31etc,PLB541,PLB563,PLB578}.

However, as for the behavior of the $U_A(1)$ anomaly in the nuclear
medium, 
the present exploratory level is rather poor. Although some theoretical
results have been reported, there exists no experimental information on
the possible effective restoration of the $U_A(1)$ anomaly at finite
density. T.~Kunihiro studied the effects of the $U_A(1)$ anomaly on
$\eta'$ properties at finite temperature using the
Nambu-Jona-Lasinio 
model~\cite{PLB219} with the KMT term~\cite{PTP44etc,PRD14etc}, which
accounts for the $U_A(1)$ anomaly effect, and showed the possible character
changes of $\eta'$ at $T\ne 0$.
There is another theoretical work with a linear $\sigma$ model~\cite{PRD29}.
Theoretical predictions by
other authors also reported the similar
consequences~\cite{PRC63,PLB560etc} and
supported the possible change of the $\eta'$ properties at
finite density as well as at finite temperature.

In this paper, we propose the formation reaction of the $\eta'$-mesic
nuclei and discuss the possibility to produce the $\eta'$-nucleus bound
states in order to investigate the $\eta'$ properties, especially mass
shift,  at finite density. 
Since the huge $\eta'$ mass is believed to have very close connection to
the $U_A(1)$ anomaly, the $\eta'$ mass in the medium should provide
us important information on the effective restoration of the
$U_A(1)$ symmetry in the nuclear medium.

\begin{figure}
\includegraphics[width=8cm]{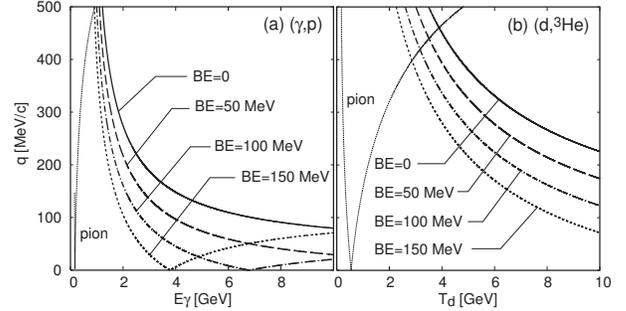}
\caption{\label{fig:mom_trans}
Momentum transfer as functions of incident particle energies for the (a)
 ($\gamma$,p) and (b) (d,$^3$He) reactions. 
Each line indicates the momentum transfer corresponding to the
 $\eta'$-mesic nucleus formation with different binding energy as shown
 in the figure. As for comparison, the momentum transfer for the pionic
 atom formation case is also shown.
}
\end{figure}
\begin{figure}
\includegraphics[width=8cm]{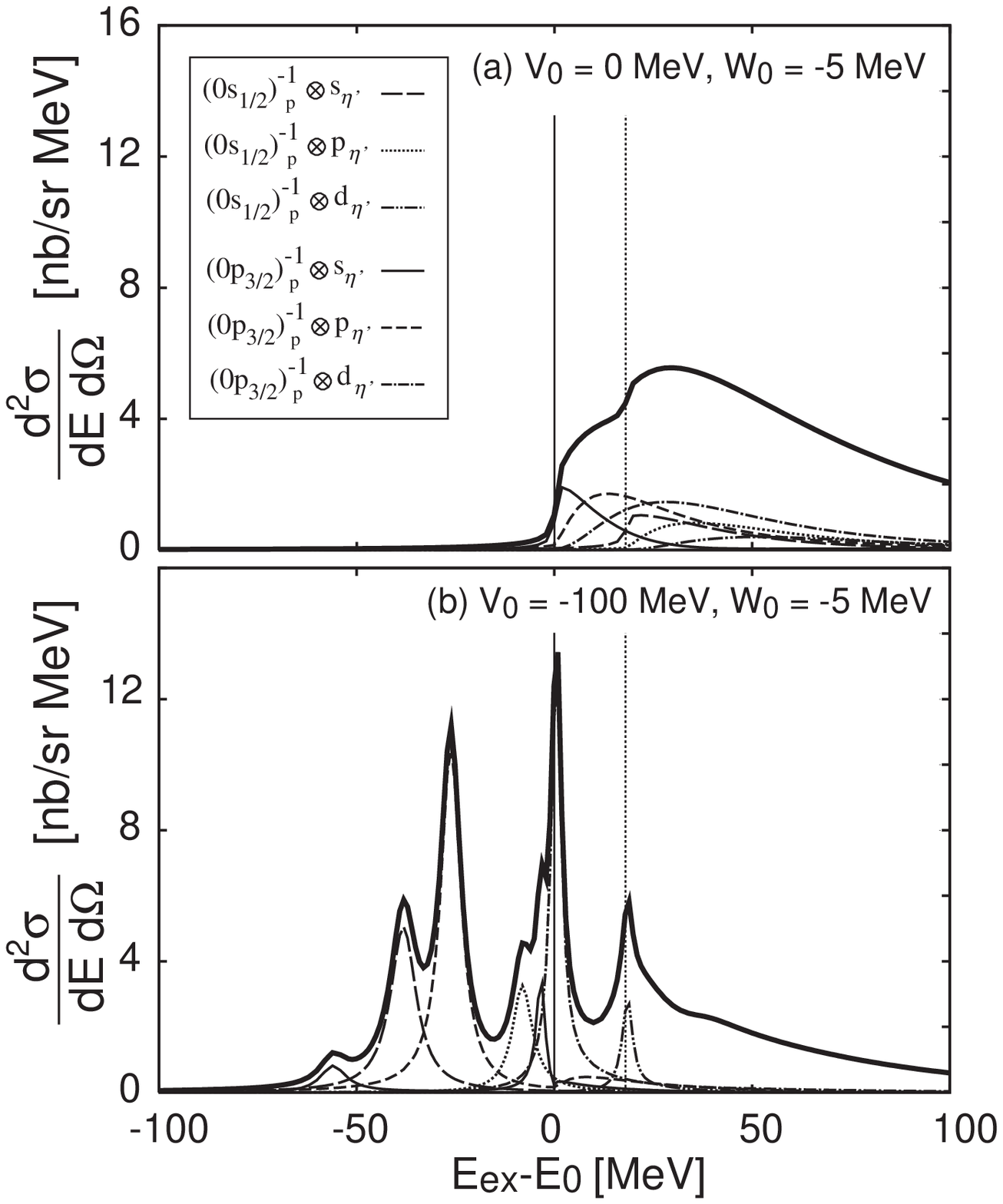}
\caption{\label{fig:W5}
The calculated spectra of $^{12}$C($\gamma$,p)$^{11}$B$\otimes \eta'$
 reaction at $E_\gamma=3$ GeV are shown as functions of the excited
 energy $E_{ex}$ defined in the text. $E_0$ is the $\eta'$ production
 threshold energy. The $\eta'$-nucleus optical potential are 
(a) $V_0=0$, $W_0=-5$ MeV and (b) $V_0=-100$ MeV, $W_0=-5$ MeV.
The total spectra are shown by the thick solid lines, and the dominant
 contributions of subcomponents are shown by dotted and dashed lines,
 as indicated in the figure.
The vertical lines indicate the $\eta'$ production threshold energy with
 the ground $p_{3/2}$ proton-hole configuration (solid line) and the
 excited $s_{1/2}$ proton-hole configuration (dotted line) in the final
 states. 
}
\end{figure}
\begin{figure}
\includegraphics[width=8cm]{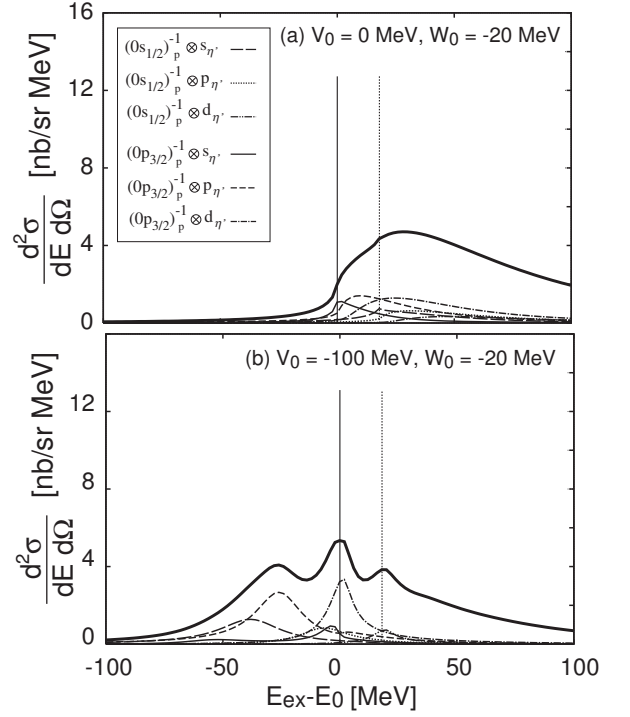}
\caption{\label{fig:W20}
The calculated spectra of $^{12}$C($\gamma$,p)$^{11}$B$\otimes \eta'$
 reaction at $E_\gamma=3$ GeV are shown as functions of the excited
 energy $E_{ex}$ defined in the text. $E_0$ is the $\eta'$ production
 threshold energy. The $\eta'$-nucleus optical potential are 
(a) $V_0=0$, $W_0=-20$ MeV and (b) $V_0=-100$ MeV, $W_0=-20$ MeV.
The total spectra are shown by the thick solid lines, and the dominant
 contributions of subcomponents are shown by dotted and dashed lines as
 indicated in the figure. 
The vertical lines indicate the $\eta'$ production threshold energy with
 the ground $p_{3/2}$ proton-hole configuration (solid line) and the
 excited $s_{1/2}$ proton-hole configuration (dotted line) in the final
 states. }
\end{figure}

In this study, we consider missing mass spectroscopy,
which was proved to be a powerful tool for the meson bound states
formation in the studies of deeply bound pionic states.
In this
spectroscopy, one observes only an emitted particle in a final state, and
obtains the double differential cross section $d^2\sigma/d\Omega/dE$ as a
function of the emitted particle energy. In order to consider
appropriate reaction for this system, we show momentum transfers as
functions of 
incident particle energies for the ($\gamma$,p) and (d,$^3$He) reactions
in Fig.~\ref{fig:mom_trans}.
The (d,$^3$He) reaction has been used experimentally for the deeply bound
pionic states formation~\cite{PRL92,PRC62etc,PRL88,PRC44}, and the
($\gamma$,p) reaction was 
proposed 
theoretically for the meson bound states
formation~\cite{PLB502,PLB527,our_gamma_p}. 
As we can see from the figure, because of
the large $\eta'$ mass, 
we need to have larger incident energies than other meson formation
cases to reduce the momentum transfer so as to have larger production
rates. We think that the ($\gamma$,p) reaction with GeV photon beam is
the appropriate reaction for our purpose since it can be performed in
existing facilities like SPring-8.
We adopt the ($\gamma$,p) reaction as a suitable one for 
the $\eta'$-mesic nuclei formation.

We choose the incident photon energy as $E_\gamma=3$ GeV, which is the
beam energy accessible at SPring-8,
and choose $^{12}$C as a target nucleus. We use the Green
function method to calculate the formation cross
sections~\cite{NPA435} as,
\begin{equation}
\left(\frac{d^2\sigma}{d\Omega dE}
\right)_{A(\gamma,p){\eta'}\otimes(A-1)}
=\left(\frac{d\sigma}{d\Omega}
\right)^{\rm Lab}_{p(\gamma,p)\eta'}
\times
\sum S(E),
\label{eq:ele}
\end{equation}
where $S(E)$ is the nuclear response function and
$\left(\frac{d\sigma}{d\Omega}\right)^{\rm Lab}_{p(\gamma,p)\eta'}$ is
the elementary cross section in the laboratory frame, which is estimated
to be 150 nb/sr using the data of SAPHIR collaboration~\cite{PLB444} and
its analysis~\cite{PRC68}. We sum up all
(proton-hole)$\otimes$($\eta'$-particle) configurations to get the total
cross section in Eq.~(\ref{eq:ele}).

To calculate the response function $S(E)$, we use the Green function
$G(E;\bm{r},\bm{r'})$ defined as~\cite{NPA435},
\begin{equation}
G(E;\bm{r},\bm{r'})=\langle p^{-1}|\phi_{\eta'}(\bm{r})
\frac{1}{E-H_{\eta'}+i\epsilon}\phi_{\eta'}^\dag(\bm{r'})|p^{-1}\rangle
\label{eq:Green},
\end{equation}
where $\phi_{\eta'}^\dag$ is the $\eta'$ creation operator and
$|p^{-1}\rangle$ is a proton hole state. The Hamiltonian $H_{\eta'}$
contains the $\eta'$-nucleus optical potential $U$. We can rewrite
Eq.~(\ref{eq:Green}) in a simple expression as,
\begin{gather}
G(E;\bm{r},\bm{r'})=\sum_{l_{\eta'},m_{\eta'}}
Y^*_{l_{\eta'},m_{\eta'}}(\hat{r})
Y_{l_{\eta'},m_{\eta'}}(\hat{r'})
G_{l_{\eta'}}(E;r,r')\\
G_{l_{\eta'}}(E;r,r')=-2m_{\eta'} ku_{l_{\eta'}}(k,r_<)v_{l_{\eta'}}^{(+)}(k,r_>), 
\end{gather}
where $u_{l_{\eta'}}$ and $v_{l_{\eta'}}^{(+)}$ respectively are the
radial part of the regular and outgoing solutions of equation of
motion. Using the Green function, the response can be calculated as
\begin{multline}
S(E)=
-\frac{1}{\pi}Im\sum_{M,m_s}\\
\int d^3rd\sigma d^3r' d\sigma'
f^\dag(\bm{r},\sigma)G(E;r,r')f(\bm{r'},\sigma).
\end{multline}
We define $f(\bm{r},\sigma)$ as
\begin{equation}
f(\bm{r},\sigma)=\chi_f^*(\bm{r})\xi^*_{\frac{1}{2},m_s}(\sigma)
\left[
Y^*_{l_{\eta'}}(\hat{r})\otimes
\psi_{j_p}(\bm{r},\sigma)
\right]_{JM}\chi_i(\bm{r}),
\end{equation}
where $\chi_i$ and $\chi_f$ respectively denote the projectile and the
ejectile distorted waves, $\psi$ is the proton hole wavefunction and
$\xi$ is the spin wavefunction introduced to count possible spin
directions of the proton in the target nucleus. 
The numerical values of $S(E)$ were evaluated by using the eikonal
approximation as in Ref.~\cite{EPJA6}.

The $\eta'$-nucleus optical potential $V(r)$ 
is assumed to have the following form as,
\begin{equation}
U(r)=(V_0+iW_0) \frac{\rho(r)}{\rho_0},
\end{equation}
where $\rho(r)$ is the nuclear density distribution and $\rho_0$ denotes
the nuclear saturation density. We treat $V_0$ as a parameter and
estimate its reasonable running range using the theoretical evaluation
of the $\eta'$ mass shift at
$\rho_0$ as $V_0=0\sim-150$ MeV~\cite{PLB219,PRC63,PLB560etc}.
We estimate the imaginary strength $W_0$ from analysis of $\gamma
p\rightarrow \eta'p$ data~\cite{nucl-th0303044}.
Since they included only $N^*(1535)$ as a baryon resonance in the
analysis of the $\eta'$ formation reaction and determined
$\eta'NN^*(1535)$ coupling strength,
we can easily
calculate the $\eta'$ 
self-energy in the medium in analogy with the $\Delta$-hole model for
the $\pi$-nucleus system as,
\begin{eqnarray}
U &\sim& \frac{g^2}{2m_{\eta'}}\frac{\rho}{m_{\eta'}+M_N-M_{N^*}+i\Gamma_{N^*}/2}\nonumber\\
&=&
(+77-8i)\frac{\rho}{\rho_0}\ {\rm [MeV]}.
\label{eq:2m0U}
\end{eqnarray}
We use the values as $-5$ MeV and $-20$ MeV for the imaginary part $W_0$
based on this evaluation in Eq.~(\ref{eq:2m0U}).
We should mention here that the evaluation in Eq.~(\ref{eq:2m0U}) provides
the repulsive real part which is opposite to the evaluation from the
$\eta'$ mass shift. If the real potential is repulsive, we do not have
any peak structure in the ($\gamma$,p) spectra due to the bound state
formation. By the ($\gamma$,p)
experiments proposed in this paper, we can
expect to distinguish these potentials and to determine the sign
and strength of the $\eta'$-nucleus optical potential.

In Fig.~\ref{fig:W5} and \ref{fig:W20}, we show the calculated spectra
as functions of the excited energy which are defined as,
\begin{equation}
E_{ex}=m_{\eta'}-B_{\eta'}+[S_p(j_p)-S_p(p_{3/2})],
\end{equation}
where $B_{\eta'}$ is the $\eta'$ binding energy and $S_p$ the proton
separation energy. The $\eta'$ production threshold energy $E_0$ is
indicated in the figure by the vertical solid lines.

We calculate four cases with
$V_0=0$ and $W_0=-5$ MeV in Fig.~\ref{fig:W5}(a), 
$V_0=-100$ MeV and $W_0=-5$ MeV in Fig.~\ref{fig:W5}(b), 
$V_0=0$ and $W_0=-20$ MeV in Fig.~\ref{fig:W20}(a), 
and $V_0=-100$ MeV and $W_0=-20$ MeV in Fig.~\ref{fig:W20}(b), 
in order to simulate 
the sensitivities of the reaction spectra to the complex potential
strength within the reasonable parameter range discussed above.

As we can see from these figures, we can expect to observe the peak
structure in the spectra due to the formation of the $\eta'$-mesic
nucleus
even in the case with the strong imaginary potential (Fig.~\ref{fig:W20}),
and we can expect to deduce the magnitude of the $\eta'$ mass
shift at finite nuclear density from the observed spectra. 
The evaluated imaginary part of the $\eta'$-nucleus potential is small
enough 
and the resonance peaks are expected to be clearly separated each
other. The absolute magnitude of the formation cross section is
reasonably large and the spectra expected be observed in experiments at
SPring-8~\cite{mura}.

The present evaluation is the first theoretical results for the
formation of the $\eta'$-mesic nuclei to know the behavior of $U_A(1)$
anomaly in the medium. We believe that the present theoretical results
is much important to stimulate both theoretical and experimental
activities to study the $U_A(1)$ anomaly at finite density and to obtain
the deeper 
insights of QCD symmetry breaking pattern and the meson mass spectrum.

We would like to thank for
D.~Jido, T.~Hatsuda, A.~Hosaka, T.~Kunihiro, M.~Oka, and M.~Takizawa for
useful comments and discussions.

\end{document}